\documentclass[useAMS,usenatbib,usegraphicx]{mn2e}

\newcommand{\hompc}{\,h\,{\rm Mpc}^{-1}}
\newcommand{\mpcoh}{\,h^{-1}\,{\rm Mpc}}
\newcommand{\Ok}{\Omega_{\rm k}}
\newcommand{\Om}{\Omega_{\rm m}}
\newcommand{\wa}{w_{\rm a}}

\usepackage{verbatim}
\usepackage{url}
\usepackage{hyperref}

\title[Cosmological model effects on redshift surveys]
{Effects of cosmological model assumptions on galaxy redshift survey measurements}

\author[Lado Samushia et al.]{
  \parbox{\textwidth}{
    Lado Samushia$^{1,2}$\thanks{e-mail: lado.samushia@port.ac.uk}, 
    Will J. Percival$^{1}$,
    Luigi Guzzo$^{3}$,
    Yun Wang$^{4}$,
    Andrea Cimatti$^{5}$,
    Carlton Baugh$^{6}$,
    James E. Geach$^{6}$,
    Cedric Lacey$^{6}$,
    Elisabetta Majerotto$^{3}$,
    Pia Mukherjee$^{7}$,
    Alvaro Orsi$^{6}$.
  }
  \vspace*{4pt} \\
  $^{1}$ Institute of Cosmology and Gravitation, University of
  Portsmouth, Dennis Sciama building, Portsmouth, P01 3FX, UK\\
  $^2$ National Abastumani Astrophysical Observatory, Ilia State University, 2A
  Kazbegi Ave, GE-0160 Tbilisi, Georgia\\
  $^3$  INAF-Osservatorio Astronomico di Brera, Via Emilio Bianchi, 46, I-23807,
  Merate (LC), Italy\\
  $^4$ Department of Physics \& Astronomy University of Oklahoma,
  Norman, OK 73019, USA\\
  $^5$ Dipartimento di Astronomia, Alma Mater Studiorum - Universit\`{a} di
  Bologna, Via Ranzani 1, I-40127, Italy\\
  $^6$ Institute for Computational Cosmology, Physics Department, Durham
  University, South Road, Durham, DH1 3LE, UK\\
  $^7$ Department of Physics and Astronomy, University of
  Sussex, Falmer, Brighton, BN1 9QH, UK\\
}

\begin{document}

\date{\today}

\pagerange{\pageref{firstpage}--\pageref{lastpage}} \pubyear{2009}

\maketitle

\label{firstpage}

\begin{abstract}

  The clustering of galaxies observed in future redshift surveys will
  provide a wealth of cosmological information. Matching the signal at
  different redshifts constrains the dark energy driving the
  acceleration of the expansion of the Universe. In tandem with these geometrical
  constraints, redshift-space distortions (RSD) depend on the build up
  of large-scale structure. As pointed out by many authors
  measurements of these effects are intrinsically coupled. We
  investigate this link, and argue that it strongly depends on the
  cosmological assumptions adopted when analysing data. Using representative assumptions for the
  parameters of the {\em Euclid} survey in order to provide a baseline future experiment, we show how the
  derived constraints change due to different model assumptions. We
  argue that even the assumption of a Friedman-Robertson-Walker (FRW)
  space-time is sufficient to reduce the importance of the coupling to
  a significant degree. Taking this idea further, we consider how the
  data would actually be analysed and argue that we should not expect
  to be able to simultaneously constrain multiple deviations from the
  standard $\Lambda$CDM model. We therefore consider different
  possible ways in which the Universe could deviate from the
  $\Lambda$CDM model, and show how the coupling between geometrical
  constraints and structure growth affects the measurement of such
  deviations.

\end{abstract}

\begin{keywords}
cosmological parameters --- cosmology: observations --- large-scale structure of the Universe
\end{keywords}

\section{Introduction}
\label{sec:introduction}

Galaxy redshift surveys will become an increasingly important source of
cosmological information. The ongoing Sloan Digital Sky Survey (SDSS) III Baryon
Oscillation Spectroscopic survey (BOSS; \citealt{schlegel09a}) will measure
redshifts for 1.5 million Luminous Red Galaxies over 10\,000\,deg$^2$, providing
cosmic variance limited constraints out to $z\sim0.6$. The next generation of
ground-based surveys, using multi-object spectrographs on 4m class telescopes
will push this cosmic variance limit to $z\sim1.4$ (e.g. BigBOSS;
\citealt{schlegel09b}). In addition to pushing beyond these redshifts, the space based
experiments benefit from having no atmospheric contamination and a larger angular
coverage compared to the ground based surveys. The European Space Agency (ESA) {\em
Euclid} mission \citep{cimatti08,laureijs09}, currently in the definition phase,
will be able to measure redshifts for galaxies out to $z\sim2$, thus measuring
large-scale structure for the full range of redshifts over which dark energy
dominates, according to standard models. Given the precision that these surveys
will achieve, it is interesting to consider exactly how the measurements from
these surveys can be used to measure the parameters of different cosmological
models.

 Galaxies are biased tracers of the underlying matter density
  field, in that they do not form a Poisson sampling of the matter
  distribution. On small scales, the number and distribution within
  each host dark matter halo is dependent on the non-linear behaviour
  of collapsed objects. On larger scales, the dark matter haloes that
  host galaxies can themselves be biased with respect to the matter
  distribution. However, on very large scales ($k\sim0.1\hompc$), the
  ratio between galaxy and matter power-spectra is scale-independent
  and the galaxy distribution can be assumed to be a fair sample of
  matter overdensities \citep[see
  e.g.,][]{kaiser84,rees85,cole89}. While modelling the full link
  between the galaxy and matter clustering is complicated, it is
  possible to use simple fits and models to extend the range of scales
  where the two power spectra can be easily linked. Although such bias
  modelling will be important for future surveys, the accuracy with
  which errors can be forecast does not depend strongly on the exact
  form of these models and we therefore assume a scale-independent bias
  in the rest of this paper. For similar reasons, we assume a linear
  power spectrum, ignoring scale-dependent growth in the clustering of
  the matter distribution.

Given the complications of galaxy bias, future Cosmic Microwave
Background (CMB) data \citep{planck06} will render the cosmological
information available from the large-scale shape of the galaxy power
spectrum or correlation function \citep{reidetal09, percivaletal07,
  tegmarketal06, coleetal05, percivaletal02} less interesting than at
present. Instead, analyses will focus on using the galaxy distribution
as a standard ruler to measure the expansion of the Universe, and use
the anisotropy and amplitude of clustering to measure the growth of
structure within it \citep{guzzo09,wang08}.

Following the Cosmological principle, the galaxy distribution is
expected to be statistically homogeneous and isotropic and measured
correlation functions and power-spectra should be spherically
symmetric in real space. In practice, this symmetry is broken:
redshift-space distortions (RSD) are present because the measured
redshift of a galaxy is not only caused by the Hubble expansion but
also has a contribution from the comoving peculiar velocity of each
galaxy with respect to the Hubble flow. Since the peculiar velocities
of individual galaxies depend on the overdensity field, the resulting
clustering signal will be angle dependent \citep{kaiser87}. RSD have
been measured using both correlation functions and power-spectra
(e.g. \citealt{hawkinsetal03, percivaletal04, zehavietal05, guzzo09,
  cabre09}).

When we use the galaxy distribution as a standard ruler, we have to model not
only the distance to the galaxies, but also the rate at which distance changes
with redshift: if we match surveys of small regions of the Universe at different
redshifts, then we need to match both the angular size (related to the distances
to the regions), {\em and} the depths of the regions. If we get the ratio of
these two projection effects wrong, we see an anisotropic clustering pattern,
which is called the Alcock-Palczynski (AP) effect \citep{alcock79}. This effect
is partially degenerate with the RSD \citep{ballinger96,simpson09}. For a
standard ruler that is small relative to scales over which cosmological
expansion becomes important, we need the angular diameter distance $R_A(z)$ to
project each part of the survey to the correct distance, and the derivative of the
radial distance ${\rm d}r(z)/{\rm d}z\equiv R_{\rm r}(z)$ to give each segment
the correct depth \citep{blake03,seo03,hu03}.

Because measurements of geometry and RSD are correlated, the cosmological
parameter measurements will also be correlated. The choice of cosmological model
to test, which acts as a prior on the measurements also acts to correlate the
measured parameters. In fact, as we show in this paper, these two effects are
strongly coupled: the importance of the measurement correlation depends on the
model to be tested. Consequently it is important when making predictions for
future surveys to clearly set out the cosmological model selection. For example,
RSD are often parametrised by $b\sigma_8$ and $f\sigma_8$, where $b$ is the
bias, $\sigma_8$ is the rms amplitude of fluctuations in the matter field in
spheres of radius $8\mpcoh$, and $f\equiv {\rm d}\log G/{\rm d}\log a$, where
$G$ is the linear growth function (e.g.
\citealt{white08,simpson09}).\footnote{The growth function can be defined as
$G(a)=\delta_{\rm k}(a)/\delta_{\rm k}(a^*)$, where $\delta_{\rm k}$ is a $k$
mode of matter overdensity and $a^*$ is a scale factor at arbitrary time $t^*$
of normalization. $G$ is a scale-independent function of redshift only on large
scales in GR, but is also a function of $k$ on small scales and in modified
theories of gravity.} In fact, this dependence follows from certain assumptions
about the Universe, which also affect the geometrical constraints (see
Section~\ref{sec:survey_measurements}). It is therefore unphysical to make
assumptions for one measurement but not for the other.  In this paper we
consider how the choice of model affects the coupling between geometrical
constraints and RSD, moving from simple assumptions about the Universe to
specific parametrizations of different models. 

In order to demonstrate these effects we predict measurements that could result
from a possible survey configuration undertaken by the {\em Euclid} satellite. We use the baseline
parameters for this survey considered by the {\em Euclid} Assessment Team, which
we briefly describe in Section~\ref{sec:euclid}. In
Section~\ref{sec:survey_measurements} we review the measurements that can be
made from galaxy surveys, and the Fisher matrix formalism by which predictions
are usually made for galaxy surveys is introduced in Section~\ref{sec:fisher}.
Section~\ref{sec:models} discusses the models and how they affect the power
spectrum. We show how the choice of model strongly affects
predictions in Section~\ref{sec:results}. We discuss our results and conclude 
in Section~\ref{sec:conclusion}.

\section{The {\em Euclid} Galaxy Redshift Survey}  \label{sec:euclid}

In order to consider the (often hidden) effect of the cosmological model
assumption on AP and RSD measurements, we consider the baseline {\em Euclid}
spectroscopic galaxy survey as outlined in the {\em Euclid} Assessment Study
Report \citep{laureijs09}. {\em Euclid} is a proposed mission to study dark
energy through an imaging survey of galaxy shapes, exploiting galaxy weak
lensing and a spectroscopic survey of galaxy redshifts exploiting the BAO
technique for measuring cosmological evolution. In this paper we only consider
the cosmological information available from the spectroscopic component of the
mission. {\em Euclid} is currently in the definition phase with possible launch
date of 2017. While these parameters can be treated as representative for a possible survey 
that could be provided by the {\em Euclid} experiment, the baseline is expected to
evolve as the definition phase progresses and consequently these numbers may not match the final 
survey achievable by this experiment.

We assume that {\em Euclid} will provide a galaxy redshift survey over a
20\,000\,deg$^2$ sky area, and will measure redshifts for emission line galaxies
over the redshift range $0.5<z<2.0$ with the precision of $\sigma_z=0.001(1+z)$.
The number density of galaxies follows the assumption that we can obtain
redshifts for 50\% of galaxies with H-$\alpha$ emission stronger than
$4\times10^{-16}$\,erg\,s$^{-1}$\,cm$^{-2}$, following the number density
distribution described in \citet{geach10}. These galaxies are biased tracers of
the mass distribution, and we adopt the redshift dependent bias relations of
\citet{orsi10}. We fit to the power spectrum over wavenumbers $k<0.2\hompc$ for
bins with $z>1.1$. For $0.5<z<1.1$, we cut this maximum scale approximately
linearly to only fit to scales $k<0.15\hompc$ at $z=0.5$ to match the increasing
scale of non-linear structure at low redshift \citep{franzetti07}. The relative
importance of different survey parameters to the ability of
constrain cosmological models parameters will be studied in \citet{wang10}.

\section{Cosmological Measurements from Galaxy  Surveys} 
\label{sec:survey_measurements}

In this paper we will only consider cosmological measurements resulting from
using RSD and from using galaxy clustering as a standard ruler. We do not
consider additional constraints from the relative clustering amplitude on small
and large scales.

\subsection{Redshift-Space Distortions}  \label{sec:rsd}

If the galaxy separation is small compared with the distance to the galaxies,
overdensities and velocities are small, the galaxy velocity field is
irrotational, and the continuity equation holds, then we can write the
redshift-space overdensity as $\delta_g^s(k)=\delta_g(k)-\mu^2\theta(k)$, where
$\theta\equiv \nabla\cdot{\bf u}$ is the divergence of the velocity field, and
$\mu$ is the cosine of the angle between wavevector $\bf k$ and the
line-of-sight \citep{kaiser87,hamilton97,scoccimarro04}. In this case the
redshift-space power spectrum $P_{gg}^s(k)$ consists of three components
\begin{equation}  \label{eq:P_3D}
  P_{\rm gg}^{\rm s}(k)=P_{\rm gg}(k)-2\mu^2P_{\rm g\theta}(k)+\mu^4P_{\theta\theta}(k),
\end{equation}
\noindent
where a subscript $g$ refers to the galaxies and a subscript $\theta$ refers to
the velocity divergence. The power spectra can be directly related if we can
assume that the linearised Euler, continuity and Poisson equations hold in a
perturbed FRW universe, and we can write 
\begin{equation} \label{eq:thetadelta}
\theta(k)=-f\delta_{\rm mass}(k),
\end{equation}
\noindent
where $f\equiv {\rm d}\ln G/{\rm d}\ln a$ is the logarithmic derivative of the linear growth
factor $G$. The linearized relationship between velocity divergence field and the matter overdensity field follows from the continuity equation and does not depend on the cosmological
model or theory of gravity, although the numerical value of $f$ will depends on the gravity model. 
If the background metric is different from FRW this relationship might not hold. Assuming that
Eq.~(\ref{eq:thetadelta}) holds,
the three power spectra have the same shape, and there are now
only two free parameters, which can be chosen to be $b\sigma_8$ and $f\sigma_8$,
giving the amplitudes (rms in spheres of radius $8\mpcoh$) of the real-space
galaxy overdensity field and velocity divergence field.

We therefore see that, in order to write the RSD dependency in terms of
$b\sigma_8$ and $f\sigma_8$, we have already had to assume that the Universe
follows a FRW model. Using $f$ and $b$ as parameters, the measured
redshift-space galaxy-galaxy power spectrum is traditionally related to the real
space matter power spectrum through
\begin{equation} \label{eq:rsd}
  P_{\rm gg}^{\rm s}(k,\mu)=P_{\rm mm}(k)(b+f\mu^2)^2.
\end{equation}

Hereafter, for simplicity and without loss of generality, we drop $\sigma_8$
from the RSD parameters, although it should be remembered that the constraints
are dependent on the amplitude of the matter overdensity field. If we drop the
assumption of a FRW model, then we can still try to constrain these parameters,
but Eq.~(\ref{eq:rsd}) no longer holds. Eq.~(\ref{eq:rsd}) also assumes that we
know the cosmological geometry in order to estimate galaxy separations.
Obviously we cannot make this assumption as we wish to measure both the RSD and
the cosmological geometry from the measured power spectrum.

In all subsequent computations we will use linear Kaiser formula of
Eq.~(\ref{eq:rsd}) to model RSD on large scales and we will also
assume that bias is not a function of $k$. Numerical simulations show
that both these assumptions are approximations even on very large
scales; linear RSD theory does not agree with simulations \citep[see,
e.g.,][]{scoccimarro04,jennings10} and galaxy bias displays scale
dependence \citep[see, e.g.,][]{angulo07} even on the scales of
$k\sim0.1\hompc$.  When analysing real data the nonlinear effects
  in RSD and bias must be included: using Eq.~(\ref{eq:rsd}) instead
  of a more accurate scale dependent model would bias the estimates of
  cosmological parameters. We presume, however, that using linear
  theory as given by Eq.~(\ref{eq:rsd}) still gives accurate estimate
  of the Fisher matrix for the galaxy survey. This should be true as
  long as the real nonlinear power-spectrum is not significantly
  different from a linear one over the range of scales included in
  this work.

\subsection{Geometrical constraints} \label{sec:ap}

We only consider galaxy clustering on radial scales that are
sufficiently small that there is negligible cosmological evolution
across them. In this case, an angular standard ruler measures
$R_A(z)/s$, and a radial standard ruler measures $R_{\rm r}(z)/s$,
where $R_A(z)$ is the (comoving) angular diameter distance, $R_{\rm
  r}(z)$ is the derivative of the radial distance, and $s$ is the
scale of the ruler. For BAO, this scale corresponds to the comoving
sound horizon at the baryon drag epoch. If, on the other hand, we use
the full correlation function or power spectrum as our ruler, this
corresponds to the average scale of the features. Forcing the observed
ruler to be the same size in radial and angular directions gives the
Alcock-Paczynski test \citep{alcock79}. In order to simplify the
equations we drop the explicit dependence on $s$ and consider that the
geometry only depends on $R_A$ and $R_{\rm r}$: it is worth
remembering that errors presented for these parameters are actually
errors on $R_A/s$ and $R_{\rm r}/s$. The ruler scale $s$ depends
  on the cosmological model parameters. We assume that the parameters
  required will be constrained by future CMB experiments with an
  uncertainty that is far smaller than could be obtained from the
  galaxy survey observations. Consequently, this dependency does not
  affect the geometrical constraints that are recovered from the
  galaxy survey analysis, and we assume that $s$ is known perfectly.

To analyse galaxy clustering data we have to adopt a fiducial cosmological
model. The angular and radial distances in our fiducial model will be different
from real distances by the factors $\alpha_{||} = R_{\rm r}/\hat{R_{\rm r}}$
and $\alpha_{\bot} = R_A/\hat{R}_A$ where quantities without a hat are computed
in a fiducial model and hat denotes real value.\footnote{These are usually
denoted by $f_{||}$ and $f_\bot$ in previous literature. Here we change the
symbols to $\alpha_{||}$ and $\alpha_\bot$ to avoid unnecessary confusion with
the growth rate $f$.} The measured components of the wavenumber along and across
the line of sight will also be different from the real ones by the same factor.
The power spectrum in Eq.~(\ref{eq:rsd}) will acquire additional angular
dependence through the AP effect and the final measured power spectrum will be
\begin{eqnarray}  \label{eq:ggmm}
\nonumber
  P_{\rm gg}^{\rm s}(k,\mu) &=& \frac{1}{\alpha_{||}\alpha^2_{\bot}}P_{\rm mm}
    \left(\frac{k}{\alpha_{\bot}}\sqrt{1+{\mu}^2(A^{-2}-1)}\right)\\
    &\times&  \left(b+\frac{{\mu}^2f}{A^2+{\mu}^2(1-A^2)}\right)^2,
\end{eqnarray}
\noindent
where $\mu = k_{||}/k$ and $A=\alpha_{||}/\alpha_{\bot}$ \citep[for details see,
e.g.,][]{ballinger96,simpson09}. The galaxy power-spectrum
is scaled by the additional factor $\Delta V = \alpha_{\bot}^{-2}\alpha_{||}^{-1}$
because the reference cosmology under(over)estimates the survey volume by the
factor of $\Delta V$.

\subsection{Survey Fisher Matrix} \label{sec:fisher}

For the most general cosmological model we consider, we model the power spectrum
of galaxies by Eq.~(\ref{eq:ggmm}) and use the parameter set ${\bf p}^{4N} =
\left\{f(z_{\rm i}),\,b(z_{\rm i}),\, \alpha_{||}(z_{\rm
i}),\,\alpha_{\bot}(z_{\rm i})\right\}$, where $1<i<N$ and $N$ is the number of
redshift slices.
For simplicity we now drop the explicit dependence on $z_{\rm i}$.

Logarithmic derivatives of Eq.~(\ref{eq:ggmm}) with respect to ${\bf p}^{4N}$
are given by

\begin{eqnarray}
\label{eq:der1}
\frac{\partial \ln P}{\partial b}& = &\frac{2}{b+f\mu^2},\\
\label{eq:der2}
\frac{\partial \ln P}{\partial f}& = &\frac{2\mu^2}{b+f\mu^2},\\
\label{eq:der3}
\frac{\partial \ln P}{\partial \alpha_{||}}& = &-1 - 4f\mu^2(1-\mu^2)/(b+f\mu^2)\\
\nonumber
& & -\mu^2\frac{\partial \ln P}{\partial \ln k}\\
\label{eq:der4}
\frac{\partial \ln P}{\partial \alpha_{\bot}}& = &-2 +
4f\mu^2(1-\mu^2)/(b+f\mu^2)\\
\nonumber
& &-(1-\mu^2)\frac{\partial \ln P}{\partial \ln k},
\end{eqnarray} 
\noindent
and the errors on radial and angular distances are related to the errors on
$\alpha_{||}$ and $\alpha_\bot$ simply by
\begin{eqnarray}
\label{eq:logeror}
\sigma_{\alpha_{||}} &=& \frac{\sigma_{R_{\rm r}}}{R_{\rm r}},\\
\sigma{\alpha_\bot}  &=& \frac{\sigma_{R_A}}{R_A}.
\end{eqnarray}

Eqns.~(\ref{eq:der1}) and~(\ref{eq:der2}) are similar to equation~(3) in
\citet{white08}, which means that AP only changes the constraints on $b$ and $f$
through cross-correlation terms in Fisher matrix, so that the errors
marginalized over $\alpha_{||}$ and $\alpha_{\bot}$ are altered from those of
\citet{white08}, but (as expected) the unmarginalized errors are not.
Eqns.~(\ref{eq:der3}) and~(\ref{eq:der4}) have three terms; the first term comes
from the effect of AP on volume, the second term is the effect of AP on RSD, the
third term is the angular dependence of isotropic power spectrum induced by AP.

Using derivatives in Eqns.~(\ref{eq:der1}) -- (\ref{eq:der4}) we will compute a
$4N$ dimensional Fisher matrix $F^{4N}$ of cosmological parameters ${\bf
p}^{4N}$ (for details see App.~\ref{app:A}).  The inverse of $F^{4N}$ gives an optimistic estimate of how well the
cosmological parameters ${\bf p}^{4N}$ will be measured in spectroscopic
surveys.

\section{Cosmological Model Assumptions}  \label{sec:models}

For a survey divided into $N$ redshift slices, which we assume to be
independent, the inverse of the $F^{4N}$ Fisher matrix gives the estimated
covariance matrix of the $4N$ cosmological parameters ${\bf p}^{4N} = \left(f,
b, \alpha_{||}, \alpha_\bot\right)$ . The only cosmological dependence of these error
estimates is that $\theta\propto\delta$ on the scales being tested. This condition
follows from FRW (see
Section~\ref{sec:rsd}), but could also hold in other types of metric. The measurement of $\alpha_\bot$ and $\alpha_{||}$ assumes that we
know either the shape of the isotropic power spectrum or at least a position of
some easily detectable feature in the power spectrum (for example position of
the first baryon acoustic oscillation peak) from other observations. In few
years time, the {\em Planck} mission will measure the linear matter power spectrum
with very high accuracy, which will strongly anchor these geometrical
constraints.

For any given cosmological model and theory of gravity, the rate of structure
growth and the radial and angular distances at different redshifts are coupled
and can be uniquely determined from a smaller number of basic physical
parameters. The reduction in the number of parameters to be constrained
obviously results in improved measurements. We will now consider how predictions
from spectroscopic galaxy surveys improve as we tighten the cosmological model.
The first and most basic assumption is that the Universe follows a FRW metric,
which we have already shown is one of the conditions required to enable RSD to
be parametrized in the standard way. 

\subsection{Friedman-Robertson-Walker Metric} \label{sec:frw}

Functions $R_A(z)$ and $R_{\rm r}(z)$ relate the coordinate angular distance and
the redshift distance to the real physical distances at redshift $z$ and, although not always
expressible in a simple form, can be defined as such in any cosmological model
and space-time.  If, however, we assume that the Universe follows a
Friedman-Robertson-Walker (FRW) metric, the radial and angular
geometrical constraints are coupled
\begin{eqnarray}
  R_{\rm r}(z) &\equiv& 
    \frac{c}{(1+z)H(z)}
    \label{eq:frwdistr}\\
  R_A(z) &\equiv& D_A(z) = \frac{c}{H_0(1+z)}\nonumber \\
    &\times&\chi\left(H_0
    \displaystyle\int_0^{z'}\frac{dz'}{H(z')}\right),
  \label{eq:frwdista}
\end{eqnarray}
\noindent
where
\begin{equation}
  \chi (x) = \left\{ \begin{array}{ll}
    x & \mbox{ if $\Ok = 0$} \\
    \frac{1}{\sqrt{\Ok}}\sin(\sqrt{\Ok}x) & \mbox{ if $\Ok>0$}\\ 
    \frac{1}{\sqrt{-\Ok}}\sinh(\sqrt{-\Ok}x) & \mbox{ if $\Ok<0$}
  \end{array} \right\},
  \label{eq:chi}
\end{equation}
\noindent
$c$ is a speed of light, $H_0$ is a Hubble constant and $H(z)$ is a Hubble
parameter which is different in every cosmological model.

In a space-time different from FRW (for example other Bianchi Type I spaces or
Lema\^{i}tre-Tolman-Bondi models) the Eqns.~(\ref{eq:frwdistr}),
(\ref{eq:frwdista}) and the relationship between the two distances is in general
different. Eqns.~(\ref{eq:frwdistr}) and~(\ref{eq:frwdista}) shows that if we
assume FRW the measurements of $H$ and $D_A$ are coupled and provide constraints
on curvature $\Ok$.

The coupling is direct: given a set of measurements $H(z_{\rm i})$ in FRW metric
the angular distance distance can be approximated as
\begin{eqnarray}
  D_A(z_{\rm i}) &=& \frac{c}{H_0(1+z_{\rm
  i})}\nonumber \\
  &\times& \chi\left[H_0\displaystyle\sum_{j}^{j\le 
  i}\left(\frac{\Delta z_{\rm j}}{H(z_{\rm j})} + \Delta I_{\rm j}
  \right)\right],
  \label{eq:dh2}
\end{eqnarray}
\noindent
where
\begin{equation}
  \Delta I_{\rm j} = O\left\{\Delta z_{\rm
  k}\partial_{zz}\left[H^{-1}(z^*)\right]\right\}
  \label{eq:interr}
\end{equation}
\noindent
is the error induced by replacing the integral with a finite sum and second
derivative of the $H^{-1}$ is evaluated at some unknown point $z^*$ inside the
interval. The $\Delta I_{\rm j} $ terms approximate the error induced by estimating the integral by a finite sum using the ``trapezoidal rule''.

After solving Eq.~(\ref{eq:dh2}) with respect to $H^{-1}(z_{\rm k})$ we can
compute derivatives
\begin{eqnarray}
  \frac{\partial H^{-1}(z_{\rm i})}{\partial D_A(z_{\rm j})} &=& 
    \frac{1 + z_{\rm l}}{c\Delta z_{\rm i}}
    \left(\delta_{\rm l,j} - \delta_{\rm l,j-1}\right) \\
    \label{eq:dfrw1}
  \frac{\partial H^{-1}(z_{\rm i})}{\partial \Ok} &=&
    \frac{H_0^2}{6c^3\Delta z_{\rm i}}D_A^3(z_{\rm l})
    (1+z_{\rm l})^3\left(\delta_{\rm l,i} - \delta_{\rm l,i-1}\right)\\
    \label{eq:dfrw2}
  \frac{\partial H^{-1}(z_{\rm i})}{\partial H_0} &=& 0,
\end{eqnarray}
\noindent
where $\delta_{\rm i,j}$ is a Kronecker delta function,
in the limit $\Ok\rightarrow0$ and transform a $4N$ dimensional Fisher matrix
$F^{4N}$ into a $3N+1$ dimensional Fisher matrix $F^{FRW}$ of variables ${\bf
p}^{FRW} = \left\{f,\, b ,\, \alpha_\bot,\, \Ok\right\}$.

\subsection{wCDM Model of Dark Energy}  \label{sec:wcdm}

Following \citet{chevallier01}, the equation of state of time-varying dark
energy is often parametrized as
\begin{equation}  \label{eq:w0wa}
  w(z) = w_0 + \wa\frac{z}{1+z}.
\end{equation}

In this model the energy density of all matter components is
\begin{equation}  \label{eq:Ew0wa}
  E(z) = \sqrt{\Om(1+z)^3 + \Ok(1+z)^2 + (1 - \Omega_{\rm m} - \Ok)F(z)},
\end{equation}
\noindent
where 
\begin{equation}  \label{eq:Fwowa}
  F(z) = (1+z)^{3(1+w_0+\wa)}\exp{\left(-3\wa\frac{z}{1+z}\right)},
\end{equation}
\noindent
and $\Om$ is relative energy density of nonrelativistic matter.

Within the wCDM model the radial and angular distances at all redshifts are
completely determined by five cosmological parameters $w_0$, $\wa$, $\Om$, $\Ok$
and $h$. Assuming wCDM in FRW space-time, but keeping $f$ as an arbitrary
function of redshift, the $4N$ dimensional Fisher matrix $F^{4N}$ now becomes a
$2N+5$ Fisher matrix $F^w$ on parameters ${\bf p}^w = \left\{f,\, b,\, w_0,\,
\wa,\, \Om,\, \Ok, h\right\}$. We will also consider an XCDM model which is a
specific case of wCDM with $\wa=0$.

\subsection{$\Lambda$CDM Model} \label{sec:lcdm}

Most cosmological data sets are consistent with a simple ``standard''
cosmological model where dark energy is time independent cosmological constant
$\Lambda$.  $\Lambda$CDM model is a specific case of wCDM with $w_0 = -1$ and
$\wa = 0$. In $\Lambda$CDM
\begin{equation}
  E(z) = \sqrt{\Om(1+z)^3 + \Ok(1+z)^2 + (1-\Omega_{\rm
      m}-\Ok)},
  \label{eq:Elcdm}
\end{equation}
\noindent
and angular and radial distances can at any redshift can be computed from just
three cosmological parameters $h$ ,$\Om$ and $\Ok$. For $\Lambda$CDM model we
transform $F^{4N}$ into a $2N+3$ dimensional Fisher matrix $F^\Lambda$ on
cosmological parameters ${\bf p}^\Lambda=\left\{f,b , h, \Om, \Ok\right\}$ and
estimate constraints on ${\bf p}^{\Lambda}$.

\subsection{$\gamma$ Parametrization of Growth} \label{sec:gamma}

In previous subsections we did not make any assumptions about the parameters $f$
and kept them as $N$ model independent numbers. If we pick a specific
cosmological model and theory of gravity the $N$ variables $f(z_{\rm i})$ will not be
independent and can be computed from a smaller number of basic cosmological
parameters.

In most conventional cosmological models and theories of gravity the
proportionality constant between matter and velocity overdensities only depends
strongly on $\Om$ and can be approximated by
\begin{equation}
  -\frac{\delta_{\rm m}}{\theta} \equiv f = 
  \left(\frac{\Om(1+z)^3}{E(z)}\right)^\gamma,
  \label{eq:gamma}
\end{equation}
\noindent
where $E(z)$ is total energy density of all matter components normalized to
$H_0$, but in modified theories of gravity is not necessarily equal to
$H(z)/H_0$.

Treating $\gamma$ as a free parameter to be fitted, $f$, $H$ and $D_A$ at all
redshifts are functions of 6 parameters ${\bf p}^{\rm \gamma w} = \left\{\gamma,
{\bf p}^{\rm w}\right\}$ in wCDM and just 4 parameters ${\bf p}^{\rm \gamma\Lambda} =
\left\{\gamma, {\bf p}^{\Lambda}\right\}$ in $\Lambda$CDM. The Fisher matrix $F^{4N}$
can then be transformed into a Fisher matrix on ${\bf p}^{\rm \gamma w}$ and
${\bf p}^{\rm\gamma \Lambda}$.

\subsection{General Relativity} \label{sec:gr}

For wCDM family of cosmological models Eq.~(\ref{eq:gamma}) with 
\begin{equation}
  \gamma = \left\{ \begin{array}{ll}
    0.55 + 0.05(1 + w_0 + 0.5\wa) &\mbox{ if $w_0\ge -1$}\\
    0.55 + 0.02(1 + w_0 + 0.5\wa) &\mbox{ if $w_0<-1$}
  \end{array} \right\}
\end{equation}
\noindent
is found to be a very good approximation to the structure growth in GR
\citep{wang98,linder05}.  For
$\Lambda$CDM this gives $\gamma = 0.55$. If GR is the correct theory of gravity
then, $f$, $D_A$ and $H$ at every redshift can be computed from just 5 parameters
${\bf p}^{\rm wGR} = \left\{h, \Ok, \Om, w_0, \wa\right\}$ in wCDM and 3 parameters
${\bf p}^{\Lambda GR} = \left\{h, \Ok, \Om\right\}$ in $\Lambda$CDM.
 
\section{Effects of Model Assumptions on Constraints}  \label{sec:results}

We will use the Fisher matrix formalism discussed above, combined with sample parameters for a
survey that could be delivered by the {\em Euclid} experiment, to investigate how
derived cosmological constraints depend on the model assumption, combining both
geometric and structure growth information. In all subsequent computations we
will assume a fiducial $\Lambda$CDM cosmology with parameters $\Om=0.25$,
$\Omega_{\rm b}=0.05$, $\Ok=0$, $\sigma_8=0.8$ and $n_{\rm s}=1.0$.

\subsection{The effect of the geometrical model on structure growth}

\begin{flushleft}
\begin{figure}
  \includegraphics[scale=0.4]{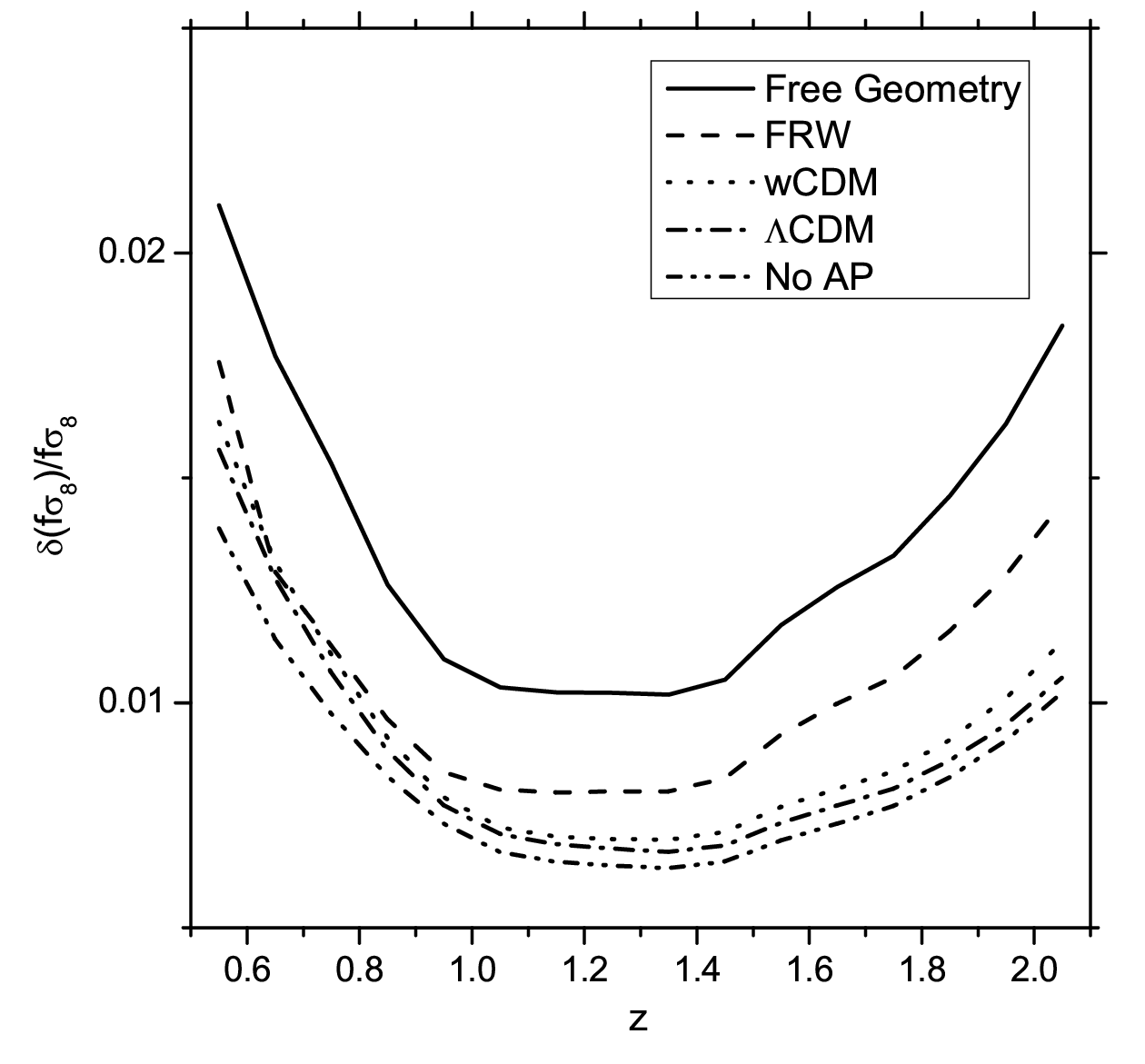}
  \caption{Constraints on $f\sigma_8$ in redshift bins of width
    $\Delta z = 0.1$ from {\em Euclid} survey with different assumptions
    about background geometry of the Universe. The solid line, labeled ``Free
    Geometry'', shows constraints on $f\sigma_8$ when no assumptions are made
    about the background cosmology.}
  \label{fig:sf}
\end{figure}
\end{flushleft}

Fig.~\ref{fig:sf} shows constraints on function $f(z)\sigma_8(z)$ in different redshift bins
for a {\em Euclid} survey for different assumptions about the model adopted for the
background geometry of the Universe.
Solid line is derived without any assumptions about background cosmology other
than the assumptions leading to Eq.~(\ref{eq:thetadelta}) in Sec.~{\ref{sec:rsd}}.
The predictions are encouraging. Even if no assumptions are made about
background cosmology {\em Euclid} can measure growth with a precision better than
2.0\$. Simply assuming FRW background brings the
constraints down to 1.5\%. As expected, the constraints get
better as we limit ourselves to models with a reduced number of basic
parameters.  Assuming wCDM or $\Lambda$CDM cosmologies further improves
constraints on $\sigma_f/f$. If we used a fixed geometry when analysing RSD the precision would be
below 1\% at intermediate redshifts. 

Fig.~\ref{fig:sf} clearly shows how big is the impact of assumptions about
geometry on the measurements of growth. We see a significant improvement even if
we only consider a FRW cosmology - similar to the assumption required to
parametrize the RSD constraints. The constraints on $f$ improve by a factor of
more than two
when we go from the most general case where we make no assumptions about
background cosmology to the best case scenario where we assume that the geometry
is known perfectly from other observations.
\begin{flushleft}
\begin{figure}
  \includegraphics[scale=0.4]{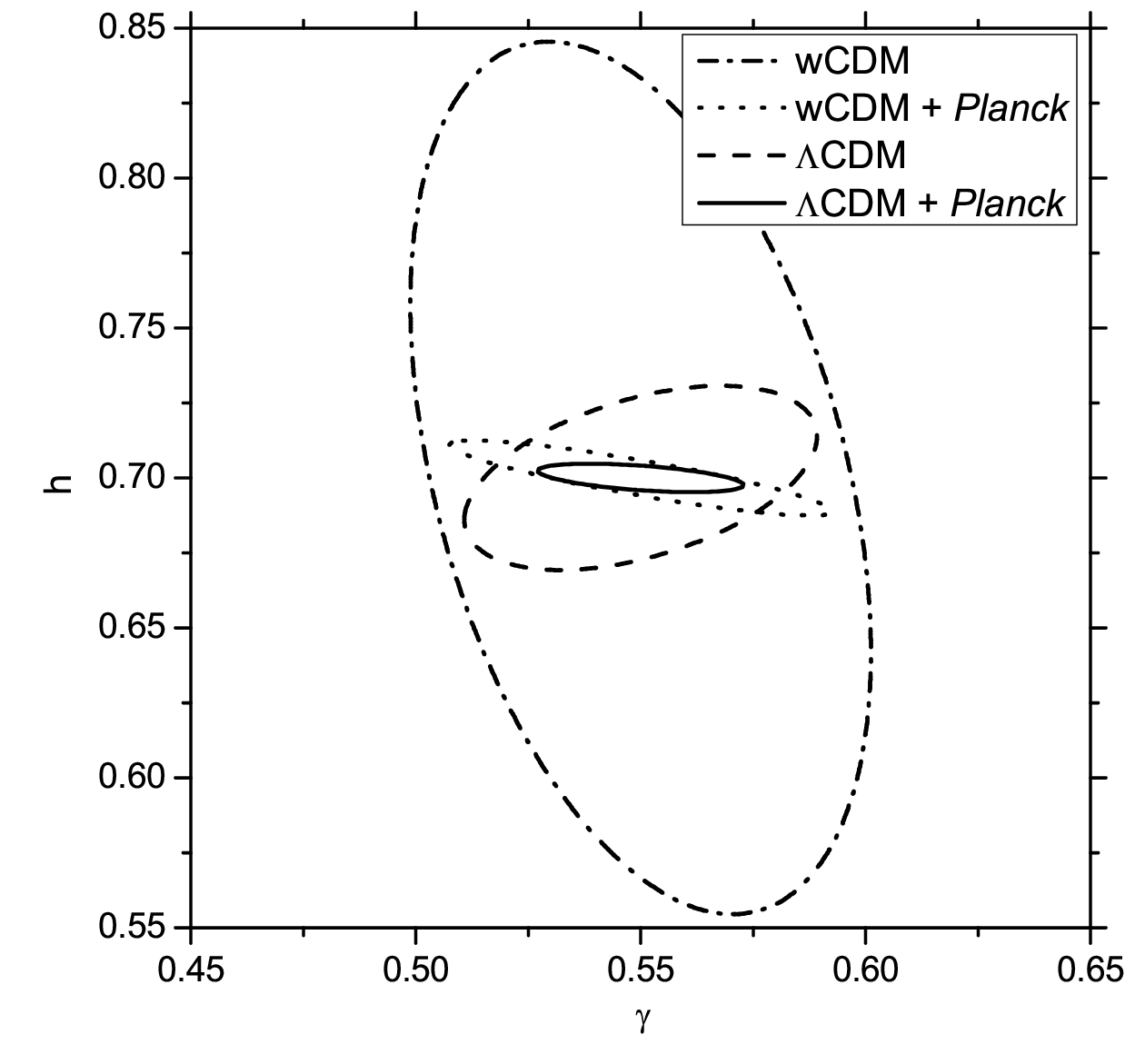}
  \caption{Constraints on parameters $\gamma$ and $h$ from {\em Euclid}
    survey with different assumptions about background cosmological
    model.}
  \label{fig:gamma}
\end{figure}
\end{flushleft}

We see similar improvements if we parametrize growth using $\gamma$ as in
Eq.~(\ref{eq:gamma}). Fig.~\ref{fig:gamma} shows that the constraints improve
significantly if we consider a $\Lambda$CDM model rather than the more general
wCDM model. Adding {\em Planck} data would make the measurements even stronger and
breaks the degeneracy between $\gamma$ and $h$ (for our treatment of {\em
Planck}
Fisher matrix see App.~\ref{app:B}). With {\em Euclid} and {\em
Planck}
measurements combined $\gamma$ can be measured to the precision of 7\% in wCDM
and to the precision of 4\% in $\Lambda$CDM, while $h$ can be measured to the
precision of 2\% in wCDM and to the precision of 1.5\% in
$\Lambda$CDM.

\subsection{The effect of structure growth assumptions on the
  geometrical model}

\begin{flushleft}
\begin{figure}
  \includegraphics[scale=0.4]{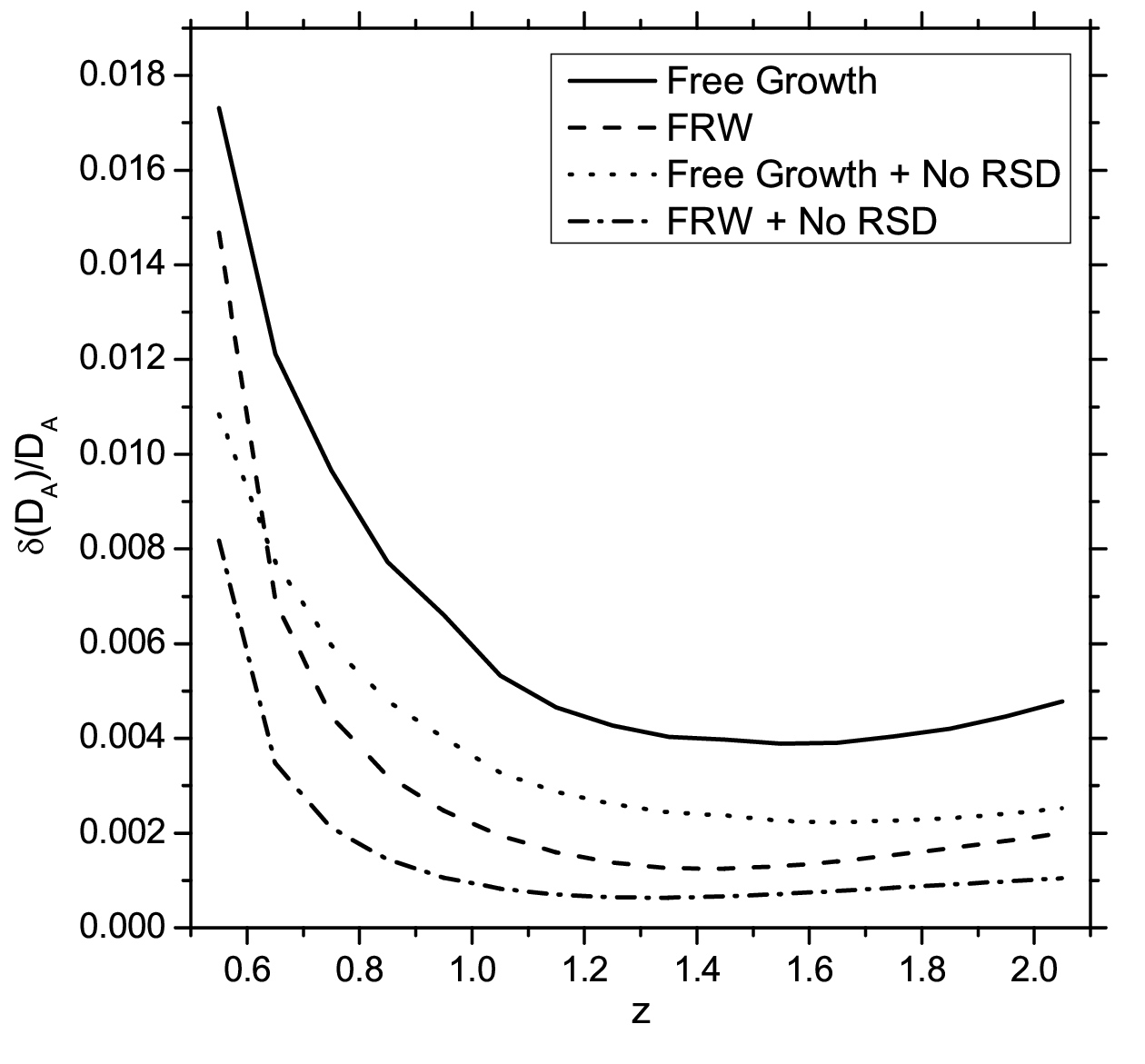}
  \caption{Constraints on angular distance $D_A$ as a function of
    redshift in redshift slices of width $\Delta z = 0.1$ for a {\em Euclid}
    survey with different assumptions about the growth of structure.}
  \label{fig:sd}
\end{figure}
\end{flushleft}

Measurements of angular and radial distances at different redshifts are strongly
affected by the assumptions about structure growth.  Fig.~\ref{fig:sd} shows how
the degeneracy with RSD affects the measurements of angular distance at
different redshifts. The solid line is derived without assuming any specific theory for growth and treating
$f(z_{\rm i})$ as independent at different redshifts and from $D_A(z)$ and $H(z)$.
If no constraints are placed on the form of the structure
growth, then geometrical constraints are degraded by a factor of $\sim$4,
compared to the case where structure growth is perfectly known. The simple
assumption of a FRW metric proves extremely significant for our ability to
measure $D_A(z)$ at intermediate redshifts as it links the angular and radial
distances through Eq.~(\ref{eq:dh2}). Adopting this assumption almost removes
the detrimental effect of having a degeneracy with unknown redshift-space
distortions effects on the geometrical constraints.
\begin{figure}
  \includegraphics[scale=0.4]{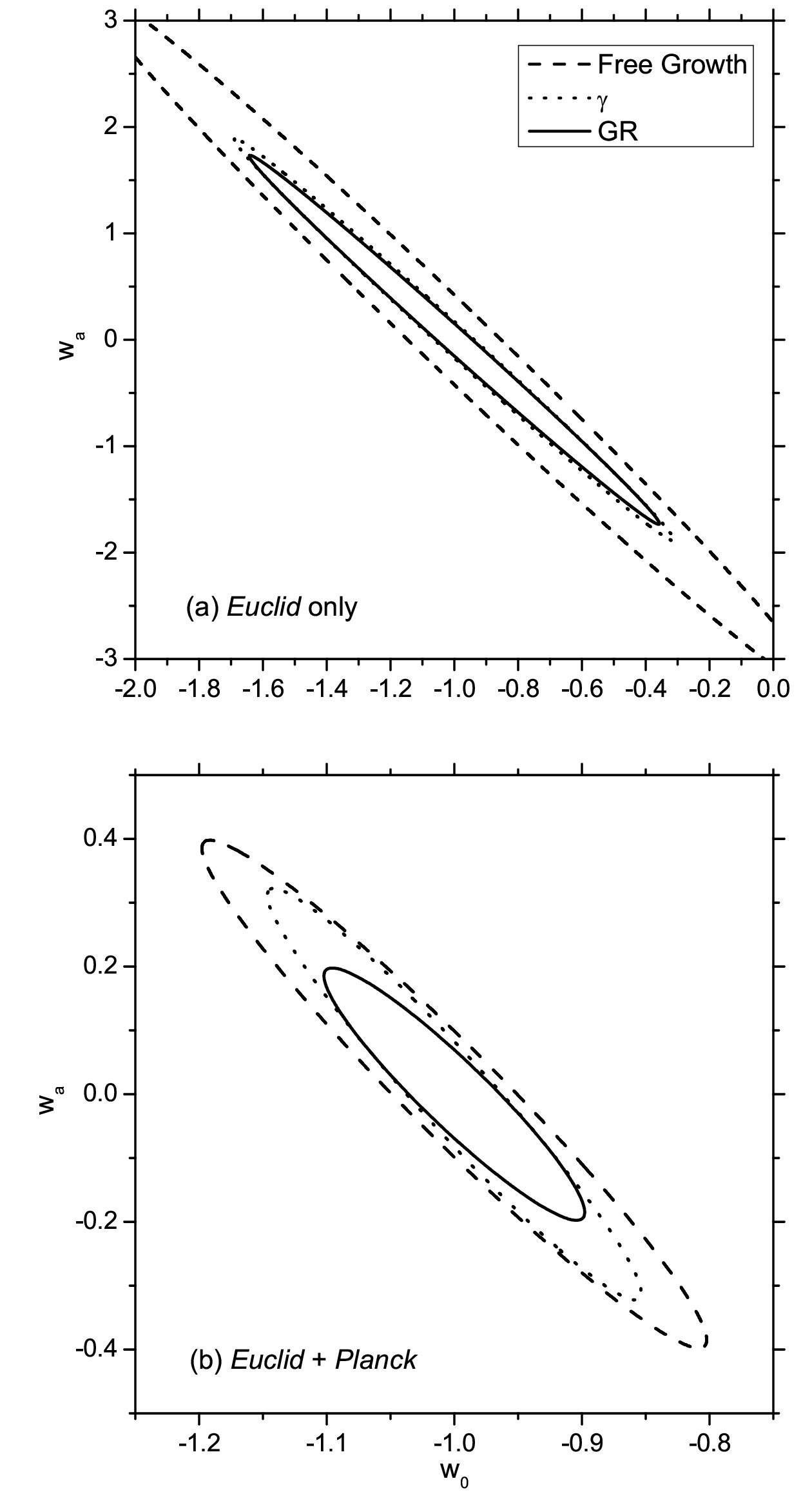}
  \caption{Constraints on cosmological parameters $w_0$ and $\wa$ from a {\em Euclid}
  survey only and from joint {\em Euclid} and {\em Planck} analyses for different assumptions about structure growth. 
  Coordinate axes on top
  and bottom panels have different scales.}
  \label{fig:w0wa}
\end{figure}

If we specify a cosmological model, the angular and radial distances at
different redshifts can be expressed in terms of smaller sets of cosmological
parameters. For the wCDM model of Section~\ref{sec:wcdm}, Fig.~\ref{fig:w0wa}
shows Fisher matrix predictions on the $w_0$, $\wa$ correlated errors when other
cosmological parameters are marginalized over. The constraints are extremely
sensitive to the assumptions about the growth of structure. If we make no
assumptions about the theory of gravity and allow the growth history to be
completely free the resulting constraints are weak, giving roughly $w_0 \in
(-1.75, -0.25)$ and $\wa \in (-2.17, 2.17)$ for marginalized errors at one $\sigma$ confidence level. When we assume that
the growth is parametrized by Eq.~(\ref{eq:gamma}) the constraints are much
stronger even when the $\gamma$ parameter is allowed to vary. The $\gamma$
parametrization results in $w_0 \in (-1.45, -0.55)$, $\wa \in (-1.25, 1.25)$
increasing the Figure of Merit (FoM) by about 4 times.\footnote{Dark Energy Task
Force (DETF)
defined the FoM as the reciprocal of the area of the error ellipse enclosing 95\%
confidence limit in $w_0$-$\wa$ plane \citep{albrecht06}.} Adding {\em Planck} priors to the {\em Euclid} measurements
results in more powerful constraints on $w_0$ and $\wa$. Assuming GR $w_0$ is now
constrained to be in the $(-1.08, -0.92)$ interval and $\wa$ is within $(-0.13,
0.13)$ at one $\sigma$ confidence level.
\begin{figure}
  \includegraphics[scale=0.4]{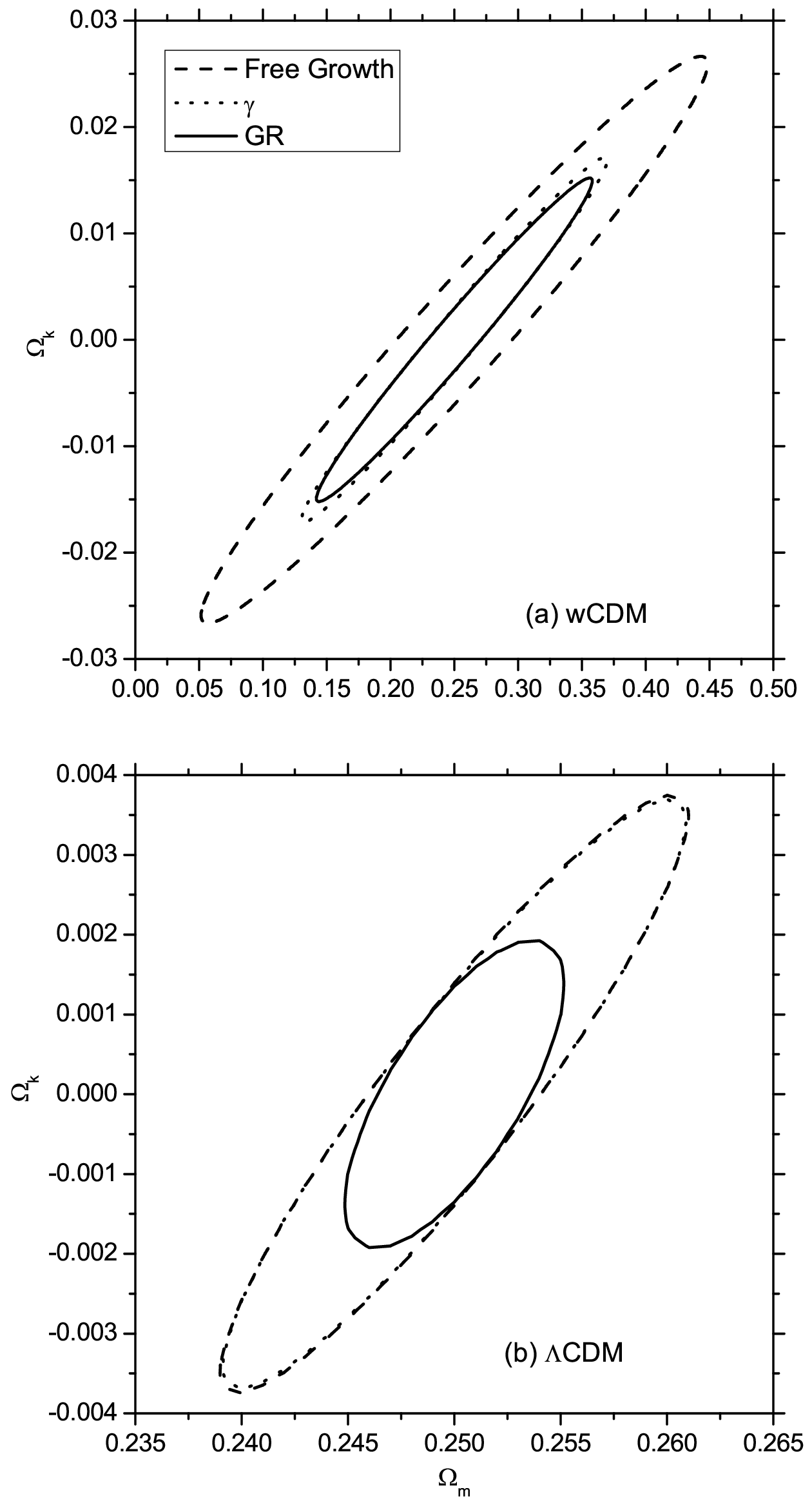}
  \caption{Constraints on cosmological parameters $\Om$ and $\Ok$ in
  wCDM and $\Lambda$CDM model from
  {\em Euclid} survey with different assumptions about the growth of structure. 
  Coordinate axes on top
  and bottom panels have different scales. The dashed and dotted lines can not
  be distinguished by eye on the bottom panel.}
  \label{fig:omok}
\end{figure}

Other cosmological parameters of interest are $\Om$ and $\Ok$. Errors on their
measurements will depend on whether we assume a time dependent dark energy
parametrized as in wCDM or time independent cosmological constant $\Lambda$.
They will also depend on the assumptions we make about growth.
Fig.~\ref{fig:omok} shows constraints on $\Om$ and $\Ok$ in wCDM and $\Lambda$CDM
scenarios with three different models for growth history. Constraints on both
parameters are extremely tight, for GR and $\Lambda$CDM the nonrelativistic matter energy density
is measured with a precision of about 1.4\% and the curvature is constrained to
be less than 0.0013 from {\em Euclid} only.

\section{Testing deviations from $\Lambda$CDM}  \label{sec:dev}

The results presented in previous sections show that the estimates on cosmological
parameters are very sensitive to the assumptions about the background geometry of
the Universe and the growth of structure. In the most general case of free
growth and unspecified geometry, the constraints on different parameters are weak
because the RSD and AP effects are degenerate. As we make stronger assumptions
about the cosmological model and theory of gravity, reducing number of
independent parameters, the degeneracy between geometry and effects of structure
growth reduces and the resulting constraints on cosmological parameters become
tight.

Because of the reasons outlined above the best method to analyse the angular
anisotropy of the measured large scale galaxy clustering data could be to fit it
to a simple ``vanilla'' $\Lambda$CDM model with GR (see first column in
Table~\ref{tab:tab1}) and then look for the
deviations from this standard model in different directions in the parameter space.

The deviations from GR are usually described in terms of the difference of
measured $\gamma$ value from its fiducial value in GR $\gamma = 0.55$ and the
deviations from cosmological constant $\Lambda$ are described in terms of
parameter $w(z)$ being different from minus one. In Table~\ref{tab:tab1} we show how
well the deviations from the simple $\Lambda$CDM and GR model can be constrained
with a future {\em Euclid} experiment.

\begin{table}
  \caption{Predicted measurements of parameters $\gamma$ and $w(z)$ around their
  fiducial value in GR and $\Lambda$CDM from {\em Euclid} experiment and {\em Euclid}
  results combined with {\em Planck} measurements.}
\begin{tabular}{|| c || c || c || c||}
  \hline \\
   & Fiducial value & 1$\sigma$ {\em Euclid} & 1$\sigma$ {\em Euclid} + {\em Planck} \\
  \hline \\
  $\gamma$    &  0.55          & 0.028        & 0.015 \\
  $w$       &  -1.00            & 0.037        & 0.0031  \\
  $w_0$     &  -1.00            & 0.42        & 0.067  \\
  $\wa$     &  0.00             & 1.14       & 0.13  \\
  \hline \\
\end{tabular}
\label{tab:tab1}
\end{table}

To get the numbers in Table~\ref{tab:tab1} we first fix a background
cosmological model to be a $\Lambda$CDM and allow the $\gamma$ parameter to
deviate from its value in GR $\gamma = 0.55$. We get a 5.0\% precision on $\gamma$
from {\em Euclid} survey and about 2.7\% precision measurement of deviation from GR
value when {\em Euclid} is combined with {\em Planck}. Then we fix the theory of gravity to
be GR and look at the deviations from the cosmological constant with $w(z) = 1$.
For the XCDM model the $w$ parameter can be constrained to be around $-1$ with a
precision of 3.7\% from {\em Euclid} and with a precision of 0.3\% with joint {\em Euclid} and
{\em Planck} analysis. In wCDM the constraints are a little looser because of the
extra parameter $\wa$. $w_0$ can be measured with a precision of 42\% around its
fiducial value with {\em Euclid} only and with a precision of 6.7\% with both {\em Euclid}
and {\em Planck}.

\section{Conclusions}  \label{sec:conclusion}

\citet{simpson09} argued that the Dark Energy Task Force \citep{albrecht06}
FoM should be expanded to include the growth of structure,
parametrised by $\gamma$ in order to allow for the degeneracy between RSD and
geometry measurements from galaxy surveys \citep{ballinger96}. However, the
importance of this degeneracy is tightly coupled with the degree of freedom
allowed in the models to be tested. We have argued that a consistent approach
needs to be adopted - any assumptions that are required to model the RSD should
also be applied to the standard ruler measurements and vice-versa. One of the
most important assumptions for the RSD follows from the assumption of a FRW
cosmology: if a FRW model is assumed when analysing structure growth then,
logically, it should also be assumed when analysing the geometry.

Care must also be taken when making predictions for future surveys to consider
how a survey will actually be analysed. Perhaps the best procedure for how to
test and constrain different cosmological models comes from the WMAP team (e.g.
\citealt{komatsuetal09}), who first fitted the ``simple'' $\Lambda$CDM model,
and then looked for deviations from this. There is a strong argument that future
galaxy surveys, such as those made possible by the {\em Euclid} satellite should be
analysed using a similar methodology. In this paper, we have argued that looking
for deviations around the baseline assumption of a $\Lambda$CDM model greatly
reduces the effect of the degeneracy between RSD and the AP effect. As the model
is relaxed and more parameters are introduced, then the degeneracy does become
more important for specific parameters. i.e. we can fit and constrain the
$\Lambda$CDM model to a high degree of accuracy, and we can find deviations
around this model, but degeneracies mean that we cannot then tell how or why
such deviations exist if we include too many degrees of freedom.

A parameter fit can be considered either as a measurement, or as a
consistency check: for example, fitting the data with a $\Lambda$CDM
model with the $\gamma$ model for structure growth can be considered a
test of General Relativity: we can test whether
$\gamma=0.55$. However, as we have seen, such tests based on the
galaxy survey data are coupled with tests of the geometrical model. In
effect, this changes the sensitivity to deviations from the
cosmological standard model to different directions. Changing the
figure of merit to be based on different parameters will simply change
the sensitivity direction. This could be chosen based on how
measurements are made (as in \citealt{simpson09}), or based on
theoretical prejudice. Here we argue that, rather than changing the
FoM, we should simply consider the most likely way in which the data
will be analysed. It seem unlikely that we will only look for
deviations by changing both the dark energy equation of state
(e.g. moving to a wCDM model), and simultaneously allowing growth of
structure to vary (e.g. moving to a $\gamma$ model for structure
growth). Instead we will look for deviations around the $\Lambda$CDM
model in particular directions and in combination. 

We provide a
C-program available at \url{http://www.icg.port.ac.uk/~samushil/Downloads/fish4d} that makes use
of publicly available GSL library, to compute a Fisher matrix and expected
errors on $f\sigma_8$ and other cosmological parameters. This should enable our
results to be checked, and constraints from both geometry and RSD to be jointly
predicted for any future survey.

\section*{Acknowledgments}
We thank Fergus Simpson for useful discussions.
LS and WJP thank the European Research Council for financial support. WJP is
also grateful for support from the UK Science and Technology Facilities Research
Council and the Leverhulme Trust.  LS acknowledges help from Georgian National
Science Foundation grant ST08/4-442 and SNSF (SCOPES grant No 128040). AC, GZ,
LG and EM acknowledge the support from the Agenzia Spaziale Italiana (ASI,
contract N. I/058/08/0). AO gratefully acknowledges an STFC Gemini studentship.

\appendix

\section{Fisher Matrix Transformations}
\label{app:A}

The Fisher matrix of cosmological parameters $\bf p$ measured from clustering
within a galaxy survey is given by
\begin{equation}  \label{eq:fisher}
  F_{\rm ij} = \frac{1}{2} \int_{k_{\rm min}}^{k_{\rm max}} 
    \frac{d^3k}{(2\pi)^3}\left(\frac{\partial \ln P}{\partial p_{\rm i}}\right)
    \left(\frac{\partial \ln P}{\partial p_{\rm j}}\right)V_{\rm eff}(k,\mu),
\end{equation}
\noindent
where the power spectrum can be measured and reliably modelled for $k_{\rm
min}<k<k_{\rm max}$. The effective volume
\begin{equation}
  V_{\rm eff} =
  \displaystyle\int{\frac{n({\bf r})P(k,\mu)}{1+n({\bf r})P(k,\mu)}d^3r =
  V_0\frac{nP(k,\mu)}{1+nP(k,\mu)}},
  \label{eq:veff}
\end{equation}
\noindent
where $V_0$ is the total volume. The second equality holds if the number density
of galaxies is constant in the volume \citep[for details see][]{tegmark97} and
the power spectrum does not significantly vary within the redshift slice. 

We compute power spectrum $P(k)$ and its derivatives for a fiducial cosmology
given by parameters in Sec.~\ref{sec:results} and
Eqns.~(\ref{eq:der1})--(\ref{eq:der4}). We use {\em Euclid} survey specifications
outlined in Sec.~\ref{sec:euclid} to compute effective volume in each redshift
shell.  The inverse of the Fisher matrix gives a covariance matrix on parameters
${\bf p}$ which to a good approximation predicts the errors on measured
cosmological parameters and correlations between them resulting from a survey in
a fiducial cosmology \citep[for details of Fisher matrix computations see,
e.g.,][]{albrecht09,bassett09}.

Using Eq.~(\ref{eq:fisher}) and derivatives in
Eqns.~(\ref{eq:der1})--(\ref{eq:der4}) we compute the initial Fisher matrix of
galaxy survey measurements as a $4N$ dimensional matrix on cosmological
parameters $f(z_{\rm i})\sigma_8$, $b(z_{\rm i})\sigma_8$, $\alpha_{||}(z_{\rm
i})$ and $\alpha_{\bot}(z_{\rm i})$. We then reduce it to the Fisher matrices of
lower dimensions by gradually imposing more restrictive assumptions about
geometry and growth. To account for the errors in distance induced by the errors
in redshift estimate we multiply the integrand in Eq.~(\ref{eq:fisher}) by a
Gaussian factor of $\exp(-k^2\Sigma_z^2)$, where $\Sigma_z^2 = \sigma_z{\rm
d}r(z)/{\rm d}z$ and $r(z)$ is the comoving distance. This has negligible
effects on our final results.

To be fully consistent we should have already included at this stage extra 
rows and columns in the Fisher matrix corresponding to the derivatives of the shape
of the power-spectrum with respect to cosmological parameters ${\bf p}$. These
elements however turn out to be very small compared to the similar elements from
Planck Fisher matrix (see, App.~\ref{app:B}) and the Fisher elements that will result
from the derivatives of growth and geometry with respect to ${\bf p}$. We can ignore
this extra information from the shape of the power-spectrum at this stage without 
significantly affecting our final results.

To go to a new set of parameters $\tilde{\bf p}$ from the old ones $\bf p$ we
use a linear transformation of a Fisher matrix
\begin{equation}
  \tilde{F}_{\rm lm} = \frac{\partial p^{\rm i}}{\partial \tilde{p}^{\rm l}}\frac{\partial
  p^{\rm k}}{\partial {\tilde p}^m}F_{\rm ij},
  \label{eq:ftrans}
\end{equation}
\noindent
with the usual summation convention over repeated indexes \citep[see,
e.g.,][]{wang06,albrecht09}.

For FRW assumption, keeping the growth and cosmological model otherwise
arbitrary, we use derivatives in Eqns.~(\ref{eq:dfrw1}) and~(\ref{eq:dfrw2}) to
get a new Fisher matrix on parameters $f(z_{\rm i})\sigma_8$, $b(z_{\rm
i})\sigma_8$, $\alpha_{\bot}$ and $\Ok$.

For wCDM model we use analytical derivatives (in the limit $\Omega_{\rm k}\rightarrow
0$
\begin{eqnarray}
  \frac{\partial \alpha_{\bot}(z)}{\partial h} &=& -\frac{1}{h},\\
  \frac{\partial \alpha_{\bot}(z)}{\partial \Ok} &=&
  \frac{1}{6\mathcal{E}(z)}\left[\displaystyle\int_0^z\frac{dz'}{E(z')}\right]^3 \nonumber
  \\
  &+& \frac{1}{2\mathcal{E}(z)}\displaystyle\int_0^z\frac{dz'}{E(z')^3}\left[F(z') -
  (1+z')^2\right],\\
  \frac{\partial \alpha_{\bot}}{\partial \Om} &=&
  \frac{1}{2\mathcal{E}(z)}\displaystyle\int_0^z\frac{dz'}{E(z')^3}\left[F(z') -
  (1+z')^3\right],\\
  \frac{\partial \alpha_{\bot}}{\partial w_0} &=&
  -\frac{1}{2\mathcal{E}(z)}\displaystyle\int_0^z\frac{dz'}{E(z')^3}\left(1-\Om-\Ok\right)\nonumber
  \\
  &\times&3F(z')\ln(1+z'),\\
  \frac{\partial \alpha_{\bot}}{\partial \wa} &=&
  -\frac{1}{2\mathcal{E}(z)}\displaystyle\int_0^z\frac{dz'}{E(z')^3}\left(1-\Om-\Ok\right)\nonumber
  \\
  &\times&3F(z')\left[\ln(1+z')-\frac{z'}{1+z'}\right],\\
  \frac{\partial \alpha_{||}(z)}{\partial h} &=& -\frac{1}{h},\\
  \frac{\partial \alpha_{||}(z)}{\partial \Ok} &=& \frac{1}{2E(z)^2}\left[F(z) - (1+z)^2\right],\\
  \frac{\partial \alpha_{||}(z)}{\partial \Om} &=& \frac{1}{2E(z)^2}\left[F(z) - (1+z)^3\right],\\
  \frac{\partial \alpha_{||}(z)}{\partial w_0} &=& -\frac{1}{2E(z)^2}(1 - \Om -
  \Ok)\nonumber \\
  &\times&3F(z)\ln(1+z),\\
  \frac{\partial \alpha_{||}(z)}{\partial \wa} &=& -\frac{1}{2E(z)^2}(1 - \Om -
  \Ok)\nonumber \\
  &\times&3F(z)\left[\ln(1+z)-\frac{z}{1+z}\right],
\end{eqnarray}
\noindent
to get a new Fisher matrix on parameters $f(z)\sigma_8$, $b(z)\sigma_8$, $h$,
$\Om$, $\Ok$, $w_0$ and $\wa$, where $F(z)$ is given by Eq.~(\ref{eq:Fwowa}) and
\begin{equation}
  \label{eq:calE}
  \mathcal{E}(z)=\displaystyle\int_0^z\frac{dz'}{E(z')}.
\end{equation}

If growth is parametrized by Eq.~(\ref{eq:gamma}), we can use measurements of
$f(z)\sigma_8$ to get constraints on $\gamma$.  We first express it as
\begin{equation}
  f(z)\sigma_8(z) = f(z)\frac{G(z)}{G(0)}\sigma_8(0),
\end{equation}
\noindent
where the growth function $G(z)$ can be expressed in terms of $\gamma$ parameter
trough
\begin{equation}
  \label{eq:Gz}
  G(z) = G(0)\exp{\left[\int_z^0{\frac{f(z')}{(1+z')}dz'}\right]}.
\end{equation}
\noindent
We than use derivatives
\begin{eqnarray}
  \frac{\partial f(z)}{\partial \gamma} &=& \frac{f(z)}{\gamma} \ln f(z),\\
  \label{eq:dg1}
  \frac{\partial f(z)}{\partial \Ok} &=& - \frac{\gamma f}{
  E(z)^2}\left[(1+z)^2 - F(z)\right],\\
  \frac{\partial f(z)}{\partial \Om} &=& \frac{\gamma f}{\Om E(z)^2
  }\nonumber\\
  &\times&\left\{E(z)^2 - \Om\left[(1+z)^3 - F(z)\right] \right\},\\
  \frac{\partial f(z)}{\partial w_0} &=& - \frac{3\gamma
  F(z)f}{E(z)^2}(1-\Om-\Ok)\ln(1+z),\\
  \frac{\partial f(z)}{\partial \wa} &=& -\frac{3\gamma
  fF(z)}{E(z)^2}(1-\Om-\Ok)\nonumber\\
  &\times&\left[\ln(1+z)-\frac{z}{1+z}\right].
  \label{eq:dg5}
\end{eqnarray}
\noindent
to transform fisher matrix elements corresponding to $f(z)\sigma_8(z)$ to the
elements of parameters, ${\bf p}^w$, $\gamma$ and $\sigma_8(0)$.

For GR we first remove the row and column corresponding to $\gamma$ parameter
and then replace it everywhere by the numerical value $\gamma=0.55$.  In
$\Lambda$CDM model we perform computations similar to wCDM case but remove the
rows and columns corresponding to parameters $w_0$ and $\wa$ and use numerical
values $w_0 = -1$, $\wa = 0$ everywhere else.

\section{{\em Planck} Fisher Matrix}
\label{app:B}

To study effects of {\em Planck} survey we utilize a {\em Planck} Fisher matrix on eight
parameters $h$, $\Om$, $\Ok$, $w_0$, $\wa$, $\sigma_8$, $n_s$ and $\Omega_{\rm
b}$ as used by DETF. 

The DETF {\em Planck } Fisher matrix is computed assuming GR and to use it with
our galaxy survey Fisher matrix we first have to generalize it
for arbitrary $\gamma\neq 0.55$. To do this we make use of the fact that CMB
experiments measure amplitude of fluctuations at the last scattering surface --
$\sigma_8(z=1100)$ -- which is related to the amplitude of density fluctuations
today -- $\sigma_8(z=0)$ -- through
\begin{equation}
  \sigma_{8,1100} = \sigma_{8,0}\frac{G_{1100}}{G_0},
\end{equation}
\noindent
where $G_z$ depends on $\gamma$ and other cosmological parameters through
Eq.~(\ref{eq:Gz}). The Fisher matrix elements of $\sigma_{8,1100}$ and
$\sigma_{8,0}$ are related
\begin{equation}
  F_{<\sigma_{8,1100};\sigma_{8,1100}>} = F_{<\sigma_{8,0};\sigma_{8,0}>}\left(\frac{G_{1100}}{G_0}\right)^2
\end{equation}
\noindent
and the Fisher matrix elements (and cross-correlation terms) on $\gamma$ are
\begin{eqnarray}
  \label{eq:Fgamma1}
  F_{<\gamma;\gamma>}&=&F_{<\sigma_{8,1100};\sigma_{8,1100}>}\left(\frac{\partial\sigma_{8,1100}}{\partial\gamma}\right)^2\nonumber\\
  &=& F_{<\sigma_{8,0};\sigma_{8,0}>}\left(\frac{G_{1100}}{G_0}\right)^2\left(\frac{\partial\sigma_{8,1100}}{\partial\gamma}\right)^2\nonumber\\
  &=& F_{<\sigma_{8,0};\sigma_{8,0}>}\sigma_{8,0}^2\gamma^2\left(\displaystyle\int_1^{1/1101}\frac{\partial f}{\partial\gamma}d\ln{a}\right)^2\\
  \label{eq:Fgamma2}
  F_{<\gamma;{\bf
  p}>}&=&F_{<\sigma_{8,1100};\sigma_{8,1100}>}\frac{\partial\sigma_{8,1100}}{\partial\gamma}\frac{\partial\sigma_{8,1100}}{\partial{\bf
  p}}\nonumber\\
  &=& F_{<\sigma_{8,0};\sigma_{8,0}>}\left(\frac{G_{1100}}{G_0}\right)^2\frac{\partial\sigma_{8,1100}}{\partial\gamma}\frac{\partial\sigma_{8,1100}}{\partial{\bf p}}\nonumber\\
  &=&
  F_{<\sigma_{8,0};\sigma_{8,0}>}\sigma_{8,0}^2\gamma\left(\displaystyle\int_1^{1/1101}\frac{\partial
  f}{\partial\gamma}d\ln{a}\right)\nonumber\\
  &\times&\left(\displaystyle\int_1^{1/1101}\frac{\partial
  f}{\partial{\bf p}}d\ln{a}\right)
\end{eqnarray}
We add a row and column corresponding to $\gamma$ to the DETF {\em Planck}
Fisher matrix and fill it with elements computed from Eqns.~(\ref{eq:Fgamma1})
and~(\ref{eq:Fgamma2}). Since our fiducial cosmology has $\gamma=0.55$ other
matrix elements do not change. The resulting 9$\times$9 matrix is a Fisher
matrix of {\em Planck} survey for a general $\gamma$. This procedure does not 
account for the fact that different value of $\gamma$ would also result in slightly different
late-time integrated Sachs-Wolf effect and would bias the estimate of $\sigma_8(0)$.
We expect, however, this effect to be small as long as $\gamma$ is within a reasonable
range ($\gamma \simeq 0.2 - 1.0$) of it's fiducial GR value. 

Before adding {\em Planck} priors we expand galaxy survey Fisher matrix rows and columns
corresponding to $f(z)\sigma_8(z)$ into rows and columns corresponding to
variables $\gamma$, ${\bf p}$ and $\sigma_{8,0}$ and add two columns padded with
zeros corresponding to
variables $n_{\rm s}$ and $\Omega_{\rm b}$. Although parameters $n_{\rm s}$ and $\Omega_{\rm b}$ can in principle be
constrained from the shape of the galaxy power spectrum
we choose not to include this information in our galaxy survey
Fisher matrix for simplicity; this is justified since the
constraints obtained from the shape of power spectrum are
significantly weaker than constraints from {\em Planck}.
 We than add the elements of nine dimensional
{\em Planck} Fisher matrix to the corresponding elements of the galaxy survey Fisher matrix.
When we work in the $\Lambda$CDM framework we remove the rows and columns
corresponding to $w_0$ and $\wa$ as before.

\bsp

\label{lastpage}

\end{document}